# Best Practices and Scoring System on Reviewing A.I.-based Medical Imaging Papers: Part 1 – Classification

Timothy L. Kline[1,*], Felipe Kitamura[2,*], Ian Pan[3], Amine M. Korchi[4,5], Neil Tenenholtz[13], Linda Moy[14], Judy Wawira Gichoya[15], Igor Santos[16], Steven Blumer[6], Misha Ysabel Hwang[7], Kim-Ann Git[8], Abishek Shroff[9], Joseph Stember[10], Elad Walach[11], George Shih[12], Steve Langer[1,+]

[1]Department of Radiology, Mayo Clinic, Rochester, MN 55905, USA
[2]DASA, Rua Gilberto Sabino, 215, São Paulo, SP 05425-020, BRAZIL
[3]Brown University, One Prospect Street, Providence, RI 02912
[4]Imaging Center Onex, Groupe 3R, 1213 Onex, Geneva, Switzerland
[5]Singularity Consulting, 1223 Cologny, Geneva, Switzerland
[6]Nemours/A.I. DuPont Hospital for Children, 1600 Rockland Road, Wilmington, DE 19803
[7]De La Salle University, 2401 Taft Avenue, Manila, Luzon 1004 Philippines
[8]Hospital Selayang, Lebuhraya Selayang-Kepong, 68100 Batu Caves, Selangor
[9]GE Healthcare, JFWTC, Bangalore, 560066 India
[10]Department of Radiology, Neuroradiology Division
Memorial Sloan Kettering Cancer Center, 1275 York Avenue, New York, NY
[11]Aidoc, Aminadav Street 3, Tel Aviv, Israel
[12]Department of Radiology, Weill Medical College of Cornell University, New York, NY
[13]Microsoft Research, Cambridge, MA
[14]Department of Radiology, NYU School of Medicine, 530 1st Ave, New York, NY 10016, USA
[15]Department of Radiology, Emory University School of Medicine
[16]Foundation Institute for Research and Study of Diagnostic Imaging (FIDI)

[+]Machine Learning Committee Sponsor

Abstract word count: 271.

Text and Abstract word count: 3,952.

**\*Shared First/Corresponding Authors:**

Timothy L. Kline, 200 First St SW, Rochester, MN 55905, 507-284-6238,
kline.timothy@mayo.edu

Felipe Kitamura, DASA, Rua Gilberto Sabino, 215, São Paulo, SP 05425-020, BRAZIL,
kitamura.felipe@gmail.com

**Running Title**: Reviewing Classification Papers

# Best Practices and Scoring System on Reviewing A.I.-based Medical Imaging Papers: Part 1 – Classification

**Abstract**

With the recent advances in A.I. methodologies and their application to medical imaging, there has been an explosion of related research programs utilizing these techniques to produce state-of-the-art classification performance. Ultimately, these research programs culminate in submission of their work for consideration in peer reviewed journals. To date, the criteria for acceptance vs. rejection is often subjective; however, reproducible science requires reproducible review. The Machine Learning Education Sub-Committee of SIIM has identified a knowledge gap and a serious need to establish guidelines for reviewing these studies. Although there have been several recent papers with this goal, this present work is written from the machine learning *practitioners* standpoint. In this series, the committee will address the best practices to be followed in an A.I.-based study and present the required sections in terms of examples and discussion of what should be included to make the studies cohesive, reproducible, accurate, and self-contained. This first entry in the series focuses on the task of image classification. Elements such as dataset curation, data pre-processing steps, defining an appropriate reference standard, data partitioning, model architecture and training are discussed.. The sections are presented as they would be detailed in a typical manuscript, with content describing the necessary information that should be included to make sure the study is of sufficient quality to be considered for publication. The goal of this series is to provide resources to not only help improve the review process for A.I.-based medical imaging papers, but to facilitate a standard for the information that is presented within all components of the research study. We hope to provide quantitative metrics in what otherwise may be a qualitative review process.

# Checklist for A.I.-based Medical Imaging Papers: Part 1 – Classification

| Section | Score | No. | Item |
|---|---|---|---|
| **Introduction** | | I-1 | A sufficient level of background information related to the relevant use case is given |
| | | I-2 | Discussion of relevant and related work is comprehensive |
| | | I-3 | Details regarding the gap the current study fills is provided |
| | | I-4 | There is a clear summary of the study objectives |
| **Methods** | | M-1 | Dataset origin is well-defined |
| | | M-2 | If data is original, appropriate oversight committee approvals are detailed |
| | | M-3 | Steps to protect patient privacy should be outlined |
| | | M-4 | Specifics regarding image acquisition parameters are clear |
| | | M-5 | Reference standard is clearly defined and justified |
| | | M-6 | The software and method for data annotation is clear and experience of those annotating is provided |
| | | M-7 | Dataset is broken up into training/validation and test sets |
| | | M-8 | The method of pre-processing is clearly defined and justified |
| | | M-9 | Data augmentation strategies are described, if used |
| | | M-10 | Discussion of the model architecture is clear |
| | | M-11 | Hyperparameter choice and training protocol are presented |
| | | M-12 | Strategies for handling class imbalance are described |
| | | M-13 | Core software versions and libraries are detailed |
| | | M-14 | Assessment of inter/intra variabilities is provided, if possible |
| | | M-15 | A primary performance metric should be identified, with appropriate rationale |
| | | M-16 | Methods for statistical analysis are described |
| | | M-17 | Sensitivity analyses were performed to test the model's robustness to various assumptions. |
| **Results** | | R-1 | A STARD-like diagram is provided |
| | | R-2 | A table with the clinical information of each data subset is shown |
| | | R-3 | Data Availability Statement is present |
| | | R-4 | Sufficient analysis is performed |
| | | R-5 | Explainability methods are used to demonstrate a reasonable and/or new explanation |
| | | R-6 | Failure analysis with reasonable hypotheses for incorrect model predictions is performed |
| **Discussion** | | D-1 | The results are summarized in a concise and coherent manner |
| | | D-2 | Proper comparison is done between the current study and the relevant ones in the recent literature. |
| | | D-3 | Limitations are thoroughly described. |
| **Conclusion** | | C-1 | Concise presentation |
| | | C-2 | Proper positioning of this work in the context of state of the art practice, if applicable |
| | | C-3 | Recommendations for future work, if applicable |
| | | C-4 | The conclusion is adequately supported by the results of the study |
| **Code** | | Co-1 | Code is made available, or if not, is justified within manuscript as to why it is omitted. |

For each item, the reviewer should provide a score in the appropriate field as follows: **0** = Item not covered, **1** = Item covered but not sufficient, **2** = Item sufficiently covered. The maximum possible score a submission can receive is 70. In general, individual items receiving a score of **0** are points warranting major revision or cause for rejection. Items receiving a score of 1 can also be highlighted in the review and likely can be easily addressed in a revision. It also may not be possible to share code, which should be considered closely by the reviewer. In general, a guide for applying these metrics is as follows: **<50** = manuscript will likely not achieve level necessary for publication even through revision and should therefore be rejected, **50-60** = manuscript needs major revision, **>60** = manuscript needs minor or no revision to be considered suitable for publication. Note also that some sections may not be appropriate for some manuscripts. Use of N/A and reduction from the total possible score should be incorporated in such cases.

**OVERVIEW**

The Machine Learning Education Subcommittee of SiiM has established this primer series with the goal of providing resources for improving manuscripts that are published in the field of medical imaging using machine learning techniques. The vision for this series is to provide a checklist for reviewers of manuscripts to help standardize the review process and identify important components that all publications should address. The first primer in this series is an overview of the key issues in study design, data curation and model training and focuses on algorithms that are developed forimage classification tasks. Future primer papers will focus in detail on other tasks such as image segmentation, as well as other areas of review (e.g., code review). We hope that this series will be useful for both the reviewers, as well as authors preparing papers in this field.

A key part of science is reproducibility. Authors must describe their methodologies and results in sufficient detail to enable readers to assess the rigor and generalizability of the work. To the extent that a work cannot be appropriately evaluated and/or reproduced, it does not advance the field. This article aims to provide a granular set of best practices to practitioners, reviewers, and editorial boards related to publishing machine learning papers related to medical image classification problems. Additional entries in the series are planned, focusing on different application areas germane to medical imaging AI. A typical classification problem in this area uses imaging data to make a prediction. For example, the prediction might try to predict tumor type, treatment response, or chances for survival. Our goal is to enhance reproducibility and replicability[1] of published manuscripts by providing a checklist and scoring rubric to guide reviewers. Other high-level guides[2–5] have been already published elsewhere and were considered when creating this guide, but in our view did not provide a quantitative way to rank a paper. We believe that this screening rubric is unique to our checklist and will highlight areas of strengths and weakness of a particular manuscript. Moreover, this scoring system may identify and confirm specific sections were there are deficiencies across multiple manuscripts. This type of analyses will allow develop strategies and resources to address these limitations and further promote the reproducibilty of AI research in medical imaging

This article is presented in a similar manner as a typical peer reviewed manuscript. Each section of this paper shows corresponding example content, followed by an expanded description for each item in the checklist. We anticipate the checklist to be used as a guide during manuscript preparation as well as review. The summary checklist serves as the guide for both reviewers and authors, and the scoring metric provides a quantifiable metric to judge submissions.

**INTRODUCTION**

The introduction section needs to motivate the study and present the specific areas that the study addresses. The use case for having the particular classification and relevant quantitative information that can be used should be clearly spelled out. References to prior studies, their successes and failures need to be comprehensive. What contributions the current study is making needs to be clear. The last paragraph should state the objective of the study.

[I-1] A sufficient level of background information related to the relevant use case is given

This includes the disease, population, and prevalence. In general, these introductory paragraphs should provide information regarding how many people are affected in a given region, what impact the disease or event has on their life (i.e. organ failure, death, need for intervention), and what is the currently accepted medical imaging modality and strategy to investigate this disease.

[I-2] Discussion of relevant and related work is comprehensive

For these paragraphs within the introduction, the authors should answer questions such as,

"How have others approached similar problems in the past?"

"What is the current state of the art in the field?"

[I-3] Details regarding the gap the current study fills is provided

Does the study describe a new approach? If so, the work should describe significant contributions beyond those proposed in prior studies.

Does the study focus on improved performance? If so, then the work should compare existing models to the proposed approach on the same data.

[I-4] There is a clear summary of the study objectives

It should be made clear the overall goals of the study and the hypothesis that led to pursuing the research. If multiple goals are pursued, the authors should identify the primary from the secondary goals.

## METHODS

The methods section should provide sufficient detail regarding data sources, approaches, and the software utilized.

[M-1] Dataset origin is well-defined

The origin of the dataset needs to be clearly detailed (original and proprietary, or public datasets). The collection and data transfer steps, in case of multi-site study, and the storage medium (on premise, cloud, internal or external) should be described. The way the quality of the dataset has been assessed should be described (human supervised or software), and finally the format of the images should be stated (DICOM, PNG, JPEG, TIFF, NIFTI, etc.). The investigators should explain the rationale of using the medical imaging exam type and specifics for US, MRI sequences, or CT phases for the study. "The process and tools used to identify, query and extract data from Electronic Medical Records/RIS/PACS should be described." The authors should provide the date range of the dataset and the specific inclusion/ exclusion criteria.

[M-2] If data is original, appropriate oversight committee approvals are detailed

The authors should report the IRB approval, and should state if locally applicable regulations (e.g., HIPAA in the USA and GDPR in Europe) have been respected. The authors should state if consent has been waived and for which reason.

[M-3] Steps to protect patient privacy should be outlined

Data privacy methods should be described (anonymization, pseudo-anonymization or de-identification) with mention to any technological tool or software used (such as encryption tools or DICOM anonymization software).

This section should also include the description of advanced anonymization techniques such as defacing on head cross sectional imaging and optical character recognition, and if human verification of each image anonymization has been performed. In addition, if metadata have been partially removed to retain useful information for training, the authors should describe which metadata have been removed and which remain available in the dataset.

## [M-4] Specifics regarding image acquisition parameters are clear

Details regarding how the images were acquired, and whether multiple vendors and/or institutional data were used should be clarified. State how data uniformity was established and the exclusion criteria to meet the protocol specifications. Provide detail regarding the data curation process to ensure wrong images were not included.
A minimum set of information should be stated for each modality:
1. Radiographs: body part and view
2. CT: body part, plane, contrast media (and bolus timing), if applicable.
3. MRI: body part, plane, sequence type, contrast media (and bolus timing), if applicable.

If data was acquired from multiples scanners/vendors, the range and distribution of acquisition parameters should be described. As protocol heterogeneity usually leads to more robust and generalizable models, it is usually a good sign to have a dataset with diverse acquisition parameters.

*"All MR acquisitions were acquired at our institution on a 'VendorName' 3T scanner ('Specific scanner details') in the supine position utilizing a multichannel surface coil. No intravenous contrast was used. The sequences included in the imaging protocol were conventional single-shot fast spin echo (SSFSE) axial, coronal and sagittal scout images followed by the T1-weighted MR images used in the study (spoiled gradient sequence, with TR=80ms, TE=3.2ms, 25º flip angle, and reconstructed voxel resolution in plane of 1.5mm, and slice thickness of 3mm). The images were acquired under a single breath-hold. A manual review guaranteed that sequences different from the ones described were not included in the dataset"*

## [M-5] Reference standard is clearly defined and justified

Details regarding the classification task and how classes are defined needs to be clear. Is this compared with pathology, surgical findings, radiological image-based descriptions, radiological reports, radiologist's impressions, etc. The reference standard is the reference against which the proposed method is compared.For a diagnostic test, this reference standard should be widely accepted test or the gold standard for the diagnosis, but it can also be based on diagnoses provided by experienced readers, especially to benchmark an algorithm designed to detect radiologic abnormalities [6].

## [M-6] The software and method for data annotation is clear and experience of those annotating is provided

The strategy and methods for annotating the data needs to be clear and its rationale explained, including:
- Markups used
- Reports used
- Level of annotation: pixel/voxel, bounding boxes, slice-level or study-level (for 3D data)
- Time tracking
- Annotation software used
- Number and experience of human annotators
- Inter/intra rater variabilities of annotations

## [M-7] Dataset is broken up into training/validation and test sets

The method for splitting the data into training, validation (development), and test sets is clearly defined. It should be clear if an independent test set was used.

Explain the dataset creation methodology and strategy (e.g., single training, validation and test split, K-fold cross-validation, etc.). State the size of each partition. Explain the rationale for the dataset size including data availability and access limitations, and statistical power estimation. If the creation of synthetic images has been used, it should be stated and the methodology explained. It is important to note the importance of using K-fold cross-validation when the dataset is small, given the large variance that can occur in performance estimation with single fold validation.

The dataset split is done at the patient level if there is more than 1 image from a specific patient in the dataset to ensure that patients are not present across multiple partitions.

## [M-8] The method of pre-processing is clearly defined and justified

Data preprocessing is critical to understand how the study can be reproduced and compared in the future. Different considerations need to be given for different modalities. For instance, CT intensities are much more standardized than those from MR. Any process that changes pixel/voxels values and numbers from the original image prior to being input to the deep learning algorithm is considered pre-processing, and such image manipulations should be specifically described. Pre-processing may include such manipulations as changing image resolution (up-sampling or down-sampling), windowing (in the case of CT), signal intensity modification and windowing (for MRI), cropping images, standardizing using the mean and/or standard deviation of image intensities, or many other image processing techniques.

*"CT Hounsfield units were windowed using a window level of 50 and window width of 100 and converted to 8-bit pixel values in the range of [0, 255]. Pixel values were then normalized to the range of [0, 1], and standardized using the normalized ImageNet mean and standard deviation per PyTorch convention. During both training and inference, images were padded to square and resized to 224x224 pixels."*

## [M-9] Data augmentation strategies are described, if used

Authors may use data augmentation methods to discourage model sensitivity to known invariants and to help improve model generalization. Authors should state which methods were performed and their corresponding parameterizations e.g. rotations of up to 15 degrees, random crops of 224x224 pixels with 0-padding as needed,etc. Some common augmentation approaches may not be justifiable in the area of medical imaging.

## [M-10] Discussion of the model architecture is clear

This includes the model architecture as well as appropriate references. Novel architectures should include a brief summary in the methods section with full details in the appendix (architecture diagrams are encouraged for clarity of presentation). State if model architecture was selected empirically or if there is a specific reason derived from known success of that architecture in the literature. Also, the use of pre-trained weights (or the weight initialization technique) should be mentioned.

*"Our model architecture is a DenseNet121 (Huang et al, 2016) backbone with 2 fully connected layers (10 classes, sigmoid activation, dropout probability 0.5) after global average pooling. Dropout probability 0.5 and sigmoid activation after global average pooling. Sigmoid activation with 10 classes was used to account for the multi-class, multi-label problem. Model was trained from scratch using He weight initialization."*

## [M-11] Hyperparameter choice and training protocol are presented

Important training and model hyperparameters (number of epochs, learning rate, model width, model depth) as well as training protocol (optimizer, loss function) are clear. If the final hyperparameters were discovered via hyperparameter optimization, the method should be detailed. Validation during training is clearly described.

*"Models were trained for 100 epochs using cross-entropy loss. The Adam optimizer with default parameters was used with an initial learning rate of $1.0 \times 10^{-3}$. Validation was performed every 5 epochs, and the learning rate was decreased by a factor of 10 if the validation loss did not improve after 10 epochs. Training was stopped early if the validation loss did not improve after 30 epochs."*

## [M-12] Strategies for handling class imbalance are described

Strategies for handling class imbalance (weighted loss, balanced sampling) are described, if used. The rationale of the different classes included in the dataset should be explained. If a specific dataset enrichment has been performed, it should be described and its rationale explained.

*"Data were sampled during training such that the proportion of each class in 1 epoch was the same across all classes."*

*"Influenza viral pneumopathy, atypical bacterial pneumopathy, and cryptogenic organizing pneumonia chest CT cases have been included in the dataset because they have similar appearances to SARS-VoV-2 viral pneumopathy, for which we are developing this model."*

[M-13] Core software versions and libraries are detailed

*"Python 3.7.6 was used with PyTorch 1.4.0 and scikit-learn 0.22"*

[M-14] Assessment of inter/intra rater variabilities is provided, if possible.

For many studies it is critical to know the inherent variability of the task, particularly for human readers. This gives the AI practitioner an idea for when a model is performing as well, or possibly better than human readers. It is also often not possible to have this information, particularly for studies that scrape data from prior reports.

[M-15] A primary performance metric should be identified, with appropriate rationale.

*"The primary metric to evaluate model performance was the area under the ROC curve, which measures the discriminatory capacity of the model."*

or

*"The primary metric of interest was the specificity at 95% sensitivity, as our model is intended to screen out negative studies without missing positive studies."[7,8]*
Authors should benchmark the performance of the AI model to radiology experts (when applicable). For classification tasks, authors should include estimates of diagnostic accuracy and their precision (such as 95% confidence intervals) (STARD 2015). When the direct calculation of confidence intervals is not possible, report nonparametric estimates from bootstrap samples (Luo 2016). Authors should apply appropriate methodology such as receiver operating characteristic (ROC) analysis and/or calibration curves. When the authors are interpreting the final model, they should report what variables were shown to be predictive of the response variable (Luo 2016).

[M-16] Methods for statistical analysis are described.

This includes presentation of the criteria for establishing statistical significance. Authors should state how diagnostic accuracy was evaluated and whether the statistical methods were appropriate for the sample size.

*"Bootstrapping was used to determine 95% confidence intervals (CI) for each performance metric and to quantify 95% CI for performance differences between radiologist and model predictions. P-values were calculated using the permutation test, p<0.05 serving as the threshold for statistical significance."*

[M-17] Sensitivity analyses were performed to test the model's robustness to various assumptions.

The authors should report on what type of sensitivity analysis was performed and whether the statistical methods were appropriate for the sample size.

*"The test set was resampled at 50%, 10%, and 1% of the original positive class prevalence to evaluate robustness against possible prevalences in the real world. To assess performance on lower quality images, images in the test set were also resized to ¼ resolution and re-interpolated to their original size. Complete results are available in supplemental materials."*

**RESULTS**

The results section should provide sufficient details to assess the performance of the approach being presented. Common ways to evaluate performance include:

- Receiver operating characteristic (ROC) curves and areas under the curve (AUCs). Often using the DeLong method or bootstrapping for statistical comparison of ROC curves.
- Decision metrics such as sensitivity, specificity, positive predictive value, and F1 score, with chosen threshold.

Discrimination performance does not guarantee good model output calibration (agreement between real probabilities and the predicted probabilities). Thus, we recommend the use of a calibration plot and/or the Hosmer-Lemeshow goodness-of-fit test[5] if well-calibrated probabilities is a desired property of the model output.

### [R-1] A STARD-like diagram is provided

A STARD-like diagram provides details regarding the number of studies/series/images before and after inclusion/exclusion criteria.

### [R-2] A table with the clinical information of each data subset is shown

A table with the clinical information of each data subset is shown, with appropriate statistical comparison. If cross-validation is performed, clinical information can be shown for the entire dataset

### [R-3] Data Availability Statement is present

Ideally, authors should try to make their datasets available. This includes the labels and the train, validation and test set indices. Note however that this may not be possible in the case of institutional data. In either case, a data availability statement should be mentioned:

> "The datasets created in this study are not publicly available because <reasons>"

> "This data is available at <repo link> under <license> license."

> "This study used an open-source dataset from <citation>."

### [R-4] Sufficient analysis is performed

This is often presented in the form of reporting accuracy, sensitivity, specificity, positive predictive value, negative predictive value, F1 score, chosen threshold, area under the ROC curve, ROC curve, area under the PR curve, PR curve are shown for test sets, when appropriate.

Appropriate statistical analysis with p-values and confidence intervals is performed.

Comparisons with radiologists are shown, if available.

Calibration is assessed, if relevant. Note that this is not always relevant, for example, it may be less important in some binary decision making problems.

Computational expense and run-times should be presented. When performed, the hardware platform should be described in detail.

[R-5] Explainability methods are used to demonstrate a reasonable and/or new explanation.

For classification tasks these are often presented in the form of saliency/attention maps used to show the image regions within the image that activate most in making a prediction/decision by the model.

[R-6] Failure analysis with reasonable hypotheses for incorrect model predictions is performed

This includes performing sanity check of incorrectly classified cases with hypothesis for failure being thoroughly described.

## DISCUSSION

The discussion should offer a concise summary of the results section, without excessive repetition. The results should be interpreted in the context of clinical utility and make suggestions for clinical use cases. Comparisons with prior studies should be performed. Limitations, especially relating to generalizability, should be stated and not overlooked.

[D-1] The results are summarized in a concise and coherent manner

This includes providing appropriate interpretation of metrics.

[D-2] Clinical use cases are suggested.

For example, a detailed discussion on how the method would change clinical workflow and what would be the benefits for the multiple stakeholders involved (patient, referring physician, radiologist, institution, etc.).

[D-3] Proper comparison is done between the current study and the relevant ones in the recent literature.

In particular, advantages and disadvantages are listed.

[D-4] Limitations are thoroughly described.

Highlighting the limitations of a study is a very important component of published literature. The authors themselves should be best placed for providing these aspects which will help future research. Important components include a discussion of what settings the results of the study are appropriate, and when they are not. For example:

Is the test data not representative of routinely acquired data?

Are there manual pre-processing steps that would be cumbersome to implement in a clinical workflow?

In general, the authors should provide guidance on what issues the reader should be aware of when interpreting the results. Some example questions that need to be presented either within the discussion section, or within other sections of the manuscript are:

Was the dataset of an appropriate size?

Were the images well standardized?

Was this multi-institution or single institution?

## CONCLUSIONS

This section should offer a "takeaway message" for the reader. As such it should focus on two areas: impact (preferably in the practical, say, clinical context) and unresolved questions that should be subjected to further research.

[C-1] Concise presentation

The conclusion does not just reiterate the sentiments provided throughout the manuscript. It should capture the main take away point(s) from the study.

[C-2] Proper positioning of this work in the context of state of the art in practice, if applicable.

How does this current work fit in with field at present?

[C-3] Recommendations for future work, if applicable.

What would the authors propose for future studies building on the current work

[C-4] The conclusion is adequately supported by the results of the study.

In general, the conclusions should not be a surprise and should be well supported by the results of the study.

## Code

[Co-1] Code is made available, or if not, is justified within manuscript as to why it is omitted

In general, review of code could be the subject of an entire paper itself. However, it is important that the details regarding code availability or lack thereof be clearly conveyed.

**Appendix**

To demonstrate the efficacy of the proposed checklist, five of the authors on this work applied the aforementioned criteria to three previously published papers for a total of 15 completed checklists. All were blinded to each other's scores until after completed evaluations were submitted. The results were then aggregated and analyzed for consistency. As the aim of this work is to formulate and assess the efficacy of the aforementioned checklist rather than evaluate previously published work, we have elected to withhold the identity of the papers used in this study.

Of the five reviewers, four provided scores that resulted in identical rankings amongst the three papers. The fifth reviewer's scores inverted the relative ordering of the first- and second- ranked papers, as deemed by the other four reviewers. Reviewer 5's difference in score between these two papers was 2 points compared to the average separation of 5.6 among all other intra-reviewer, inter-paper comparisons. Together, these results highlight the ability of the checklist to delineate between works of varying quality.

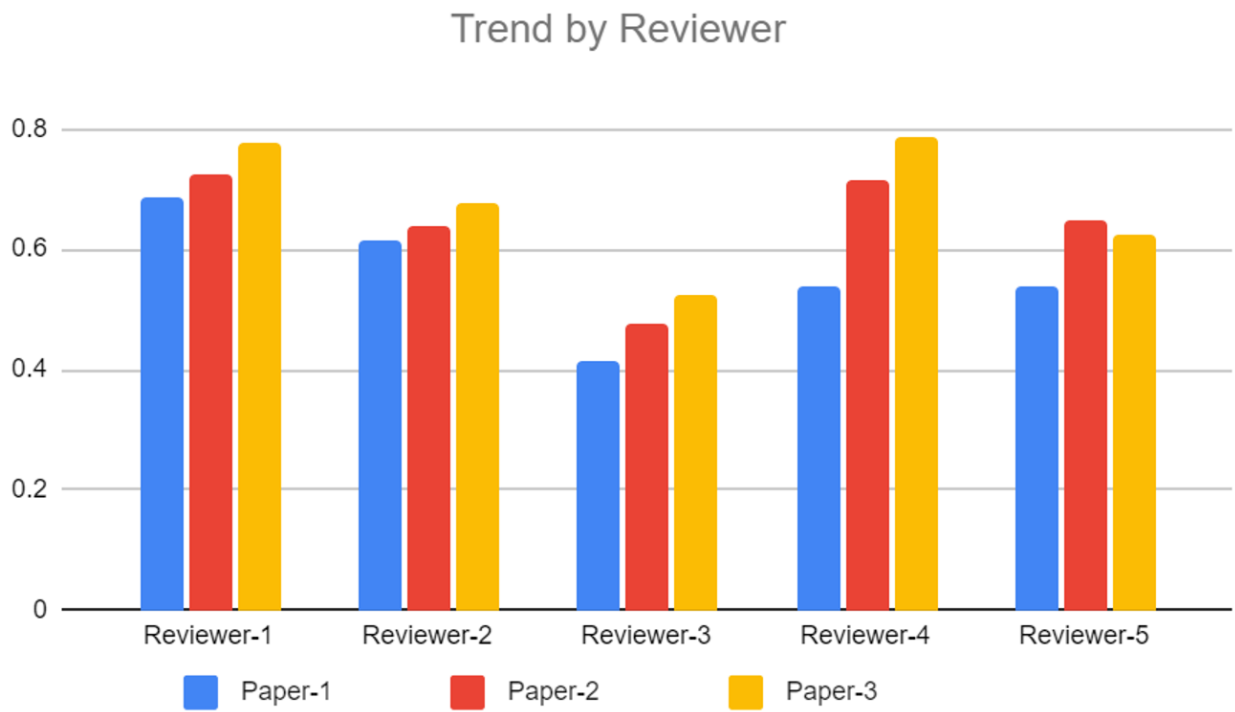

**Figure 1.** Summary of the trend by the five reviewers for the three different classification papers. The rank order of the papers was consistent, with only one reviewer rating Paper-3 lower than Paper-2.

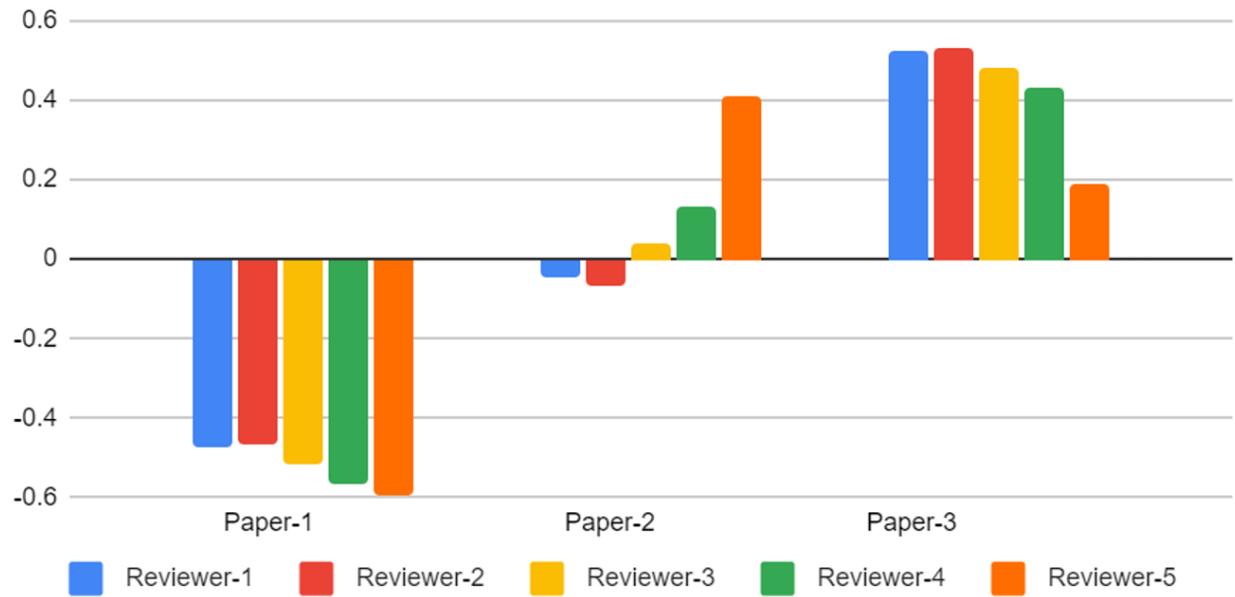

**Figure 2.** User normalized reviewer variability by paper highlighting the consistent nature of the rank order of the papers by all five reviewers.

**Table 1.** Analysis of the individual checklist sections.

| Reviewer | Section | Paper-1 | Paper-2 | Paper-3 |
|---|---|---|---|---|
| Reviewer-1 | Total | 0.69 | 0.73 | 0.78 |
| | Intro | 1.00 | 0.63 | 1.00 |
| | Methods | 0.85 | 0.82 | 0.85 |
| | Results | 0.33 | 0.75 | 0.75 |
| | Discussion | 0.75 | 1.00 | 1.00 |
| | Conclusion | 1.00 | 1.00 | 1.00 |
| | Code | 0.00 | 0.00 | 0.00 |
| Reviewer-2 | Total | 0.61 | 0.64 | 0.68 |
| | Intro | 1.00 | 1.00 | 1.00 |
| | Methods | 0.79 | 0.65 | 0.82 |
| | Results | 0.50 | 0.67 | 0.42 |
| | Discussion | 0.50 | 1.00 | 0.88 |
| | Conclusion | 0.50 | 0.63 | 0.75 |
| | Code | 0.00 | 0.00 | 0.00 |
| Reviewer-3 | Total | 0.41 | 0.48 | 0.53 |
| | Intro | 1.00 | 0.75 | 0.88 |
| | Methods | 0.44 | 0.41 | 0.44 |
| | Results | 0.08 | 0.33 | 0.33 |
| | Discussion | 0.50 | 1.00 | 1.00 |
| | Conclusion | 0.63 | 0.75 | 1.00 |
| | Code | 0.00 | 0.00 | 0.00 |
| Reviewer-4 | Total | 0.54 | 0.71 | 0.79 |
| | Intro | 0.63 | 1.00 | 1.00 |
| | Methods | 0.71 | 0.82 | 0.91 |
| | Results | 0.08 | 0.42 | 0.67 |
| | Discussion | 0.88 | 1.00 | 1.00 |
| | Conclusion | 0.75 | 1.00 | 1.00 |
| | Code | 0.00 | 0.00 | 0.00 |
| Reviewer-5 | Total | 0.54 | 0.65 | 0.63 |
| | Intro | 1.00 | 0.88 | 0.88 |
| | Methods | 0.59 | 0.79 | 0.76 |
| | Results | 0.17 | 0.33 | 0.33 |
| | Discussion | 0.88 | 1.00 | 0.63 |
| | Conclusion | 0.75 | 0.75 | 1.00 |
| | Code | 0.00 | 0.00 | 0.00 |